\documentclass[12pt,showpacs,showkeys,preprintnumbers,nofootinbib]{revtex4}

\usepackage{amssymb,amsmath,graphics,graphicx}

\begin{document}

\newcommand{\pderiv}[2]{\frac{\partial #1}{\partial #2}}
\newcommand{\deriv}[2]{\frac{d #1}{d #2}}

  \title{Information spreading with aging in heterogeneous populations}

\author{Marlon Ramos $^{1}$}
\thanks{E-mail: marlon.ramos@fis.puc-rio.br}

\author{Nuno Crokidakis $^{1}$}
\thanks{E-mail: nuno.crokidakis@fis.puc-rio.br}

\author{Celia Anteneodo $^{1,2}$}
\thanks{E-mail: celia@fis.puc-rio.br}

\affiliation{
$^{1}$Departamento de F\'{\i}sica, PUC-Rio, Rio de Janeiro, Brazil \\
$^{2}$National Institute of Science and Technology for Complex Systems, Rio de Janeiro, Brazil}

\date{\today}

\begin{abstract}
\noindent
We study the critical properties of a model of information spreading based on the SIS epidemic model. 
Spreading rates decay with time, as ruled by two parameters, $\epsilon$ and $l$, 
that can be either  constant or randomly distributed in the population.  
The spreading dynamics is developed on top of Erd\"os-Renyi networks.
We present the mean-field analytical solution of the model in its simplest formulation, 
and Monte Carlo simulations are performed for the more heterogeneous cases. 
The outcomes show that the system undergoes a nonequilibrium phase transition 
whose critical point depends on the parameters $\epsilon$ and $l$. 
In addition, we conclude that the more heterogeneous the population, 
the more favored the information spreading over the network.

\end{abstract}



\maketitle

\section{Introduction}

  In the last decades, diverse questions of social dynamics have 
been tackled by means of statistical physics techniques. 
In fact, simple models allow to simulate and understand real problems such as elections, 
spread of information, vehicle traffic or pedestrian evacuation, amongst 
many others \cite{loreto_rmp}. As a feedback, these issues are attractive to physicists 
because of the occurrence of order-disorder transitions, scaling and universality, 
among other typical features of physical systems. 

More recently, due to the emergence and popularization of social networks like facebook and twitter, 
as means of information dissemination, 
there is a growing interest in the study of rumor and information spreading in complex networks. 
For this purpose, a large diversity of models have been 
proposed \cite{dk,galam,moreno1,moreno2,vazquez,zanette,tang,zhao1,zhao2,zhao3,zhao4,moro,kun}. 
The majority of these models are based on the standard ones of epidemic spreading like SI, SIS, SIR 
and their variants \cite{anderson_may}. 
The paradigmatic model of rumor spreading is the Daley-Kendal (DK) model \cite{dk}, 
that is conceptually similar to the SIR model. 
The population is divided into three distinct states, namely Spreaders, Ignorants and Stiflers. 
The Spreaders are agents that are transmitting the rumor through the population, 
the Ignorants do not know the rumor and, consequently, they are not spreading, 
and finally the Stiflers are those individuals who know the rumor but 
have lost interest in diffusing it. 
The transitions between states are given by stochastic rules in the same way as in  
epidemic models. Many extensions of the DK model were studied by the consideration of random, 
scale-free  \cite{moreno1,moreno2,vazquez} and small-world \cite{zanette} networks 
for the contact among individuals, two different kinds of rumors spreading over 
the network \cite{tang}, new classes of individuals \cite{zhao1}, effects of media \cite{zhao2}, 
remembering and loss of memory  \cite{zhao3,zhao4},  
impact of human activities \cite{moro}, among others.

Our present motivation is to investigate some of the mechanisms involved 
in the adoption of innovations, new ideas or technologies. 
This issue may have practical applications such as in marketing strategies 
and have already been target of recent studies  \cite{kun,toole,iglesias}. 
We focus on the spreading of information of the kind that induces the adoption of a new 
product or idea. Although, for practical purposes, one has primarily in mind 
the promotion of new goods introduced in the market, our proposal may also apply, 
for instance, to political propaganda. 
For that goal we consider a dynamics of information spreading inspired 
on the SIS epidemic model. Each agent can be in one of two possible states, 
namely, $S$ (Spreader) or $R$ (Restrained). 
The individuals in state $S$ are those spreading the information through the network, 
whereas the agents in the state $R$ are not transmitting it even if they are aware 
of the information. 
Notice that for marketing purposes, it is important that the agents 
become enthusiastic transmitters and not only that 
they know or adopt the technology.  
Agents initially in state $R$ do not have the knowledge. 
When such an agent is put into contact with a new technology, 
product or ideology, by interaction with Spreaders, the agent becomes  with rate $\lambda$  
an enthusiastic adopter or 
spreader, trying in turn to convince the  neighbors in the network of contacts 
(friends, relatives, etc.) to adopt the innovation. 
We denote this state as $S$. Frequently, this enthusiasm is transient and, after some time period, 
the agent spontaneously decays to the state $R$ with rate $\alpha$, 
where even though the agent may know or use the technology, he/she does not propagate it. 
Also this state is not permanent and the agent could become again a Spreader, however, 
with a reduced transition rate. 
To take this fact into account, we propose that, once the individual decays to state $R$, 
his/her ``contagion'' rate $\lambda$ decreases as $\lambda\to \epsilon\,\lambda$, with $\epsilon<1$. 
It reflects the fact that individuals tend to become more and more resistant to spread the information. 
Nonetheless, the reduction of the ``infection'' rate occurs only a limited number $l$ of times.

Both parameters, $\epsilon$ and $l$, can be either uniform, 
i.e., equal for all individuals, or  vary from one individual to another. 
In this work we will study the effect of the aging of transition rates 
on the phase diagram of the model.  
Moreover, we will show the crucial role of 
heterogeneities on the critical behavior and its consequences for information spreading. 
%
%We found that the system undergoes 
%nonequilibrium phase transitions for all studied cases, and the transition points 
%depend on the model parameters.

The paper is organized as follows. In Section II we present the general formulation 
of the model and define its microscopic rules. 
The analytical and numerical results of four distinct cases analyzed are discussed in Section III. 
Section IV contains the conclusions and final remarks.

% ##########################################################################

\section{The model}

  We have considered a dynamics of information spreading based 
on the SIS epidemic model, where each agent can be in one of two possible states. 
In our case,   $S$ (spreader) or $R$ (restrained). 
The microscopic rules, based on a variant of  the SIS model with aging effects \cite{nuno_marcio}, 
 are the following:

\begin{enumerate}

\item Each individual $j$ in the $S$ state at time $t$ becomes $R$ 
in the next time step $t+1$ with probability $\alpha$; 

\item Each individual $j$ in the $R$ state at time $t$ becomes $S$ 
in the next time step $t+1$ with probability $\lambda_{j}(t)$ 
if it has at least one neighbor in the $S$  state;

\item Each individual $j$ starts the dynamics with $\lambda_{j}(t=0)=\lambda_{0}$. 
After each transition $S\to R$, the spreading probability $\lambda_{j}$ is updated according to
\begin{equation} \label{lambda_j}
\lambda_{j}(t+1)=\epsilon_{j}\, \lambda_{j}(t) ~,
\end{equation}
where, for all $j$, $\epsilon_{j}<1$ is the factor of reduction of the spreading rate 
(or probability of spreading within a time unit).

\item In addition, the update given by Eq. (\ref{lambda_j}) occurs 
a maximal number $l_{j}$ of times for each individual $j$.

\end{enumerate}

Let us recall that, initially, the individuals in the $R$ state do not know the information. 
Actually, this would correspond to a third state (Ignorant agent), but this 
state only occurs at the beginning of the dynamics and it is not attainable later on, as soon 
as oblivion is not taken into account. 
After a contact with an agent $S$ (spreader), an individual $j$ in state $R$ 
comes to know the information, hence becoming $S$, with probability  $\lambda_{j}$ 
(this probability is, of course, equal for all individuals at $t=0$, 
i.e., $\lambda_{j}=\lambda_{0}$ for all $j$). 
After that, the individuals will not forget the information, but they can spread it or not. 
In standard models of rumor spreading, 
the Ignorants do not spread the rumor  because they do not know it, 
while the Stiflers know the rumor but they do not transmit it. 
In comparison with the standard states considered in rumor models, 
the agents in the $R$ state can be identified with Ignorants only 
at the beginning, when they do not have acquired the knowledge yet, while they 
can be identified with Stiflers after each transition $S \to R$, in which case they do 
have the knowledge. 
Moreover, concerning the transition rules in standard models, 
Stiflers do not become Spreaders again, unlike in our model.

Notice, from our rules, that the spreading probabilities are heterogeneous, 
varying from one individual to another. 
Each agent $j$ who stops  spreading the information through the network (Restrained) may become 
a Spreader again, but the probability with which this event occurs decreases with time, 
depending on the agent intrinsic traits, given by ($\epsilon_{j}, l_{j}$). 
After each transition $S\to R$, the spreading probability of a given agent $j$ decreases, 
up to a maximal number of times $l_j$, then, in the next time step, 
it will be more difficult for the contacts of the individual $j$ 
to ``persuade'' him/her to spread the information again.

%%%%%%% motivation

We have investigated the model on top of an Erd\"os-Renyi (ER) network with size $N$ 
and Poissonian degree distribution $P(k)=e^{-\langle k\rangle} \langle k\rangle^k/k!$. 
In this case, we have computed information spreading only in the largest connected 
component, i.e., the giant cluster. 
Unless otherwise stated, we have considered networks with $\langle k\rangle=10$, 
for which the probability that a given node belongs to the giant cluster 
is approximately given by $0.99995$ in the thermodynamic limit \cite{er}.

We will analyze four distinct cases: 
(i) uniform parameters, i.e., $\epsilon_{j}=\epsilon$ and $l_{j}=l$ 
for all individuals $j$; (ii) uniform $\epsilon$ and random $l$; 
(iii) random $\epsilon$ and uniform $l$, and finally (iv) random $\epsilon$ and random $l$. 
When randomness is considered, the uniform probability distribution is used to generate the parameters.

%%%%%%%%%%%%%%%%%%%%%%%%%%%%%%%%%%%%%%%%%%%%%%%%%

\section{Results}

\subsection{Uniform $\epsilon$ and uniform $l$}

  In this case, we have $\epsilon_{j}=\epsilon$ and $l_{j}=l$ 
for each individual $j$. However, there is a certain degree of heterogeneity 
in the system due to the distinct histories of the spreading rates $\lambda_{j}$. 
Following the mean-field approach used to treat epidemic models \cite{pastor_satorras}, 
one obtains the analytical solution of the model. Initially, 
let us define $S_{m}$ as the number of Spreaders that have performed 
the transition $S\to R$ exactly $m$ times, where $m$ can take the values 
$m=0,1,...,l$. 
Observe that these agents have spreading rates $\lambda_{m}=\lambda_{0}\,\epsilon^{m}$. 
At mean-field level, one can assume that after a long time 
(but before attaining the steady state) all individuals will be in either one of two 
states %\footnote{In the following, we will verify that this assumption is valid.}, 
\cite{footnote1}, 
namely, either $S_{l}$ or $R$ \cite{nuno_marcio}. 
Thus, the only relevant equation to the time evolution of the system is
\begin{eqnarray}\label{eq2}
\dot{s_{l}} = -\alpha\,s_{l} + \lambda_{l}\langle k\rangle s_{l}\,(1-s_{l}) ~, 
\end{eqnarray}
where $s_{l}=S_{l}/N$ is the density of Spreaders in the state $S_{l}$, 
$\lambda_{l}=\lambda_{0}\,\epsilon^{l}$ and we have used the normalization condition 
$s+r=1$. 
Due to the Poisson distribution of ER graphs, in Eq. (\ref{eq2}), we have neglected the 
fluctuations on the connectivity and made the approximation that every node has the same degree 
\cite{pastor_satorras}.

In the steady state, $s_{l}=s$ and $\dot{s_{l}}=\dot{s}=0$, 
which leads either to the trivial solution $s=0$ or to 
\begin{eqnarray}\label{eq3}
s = 1 - \frac{\alpha}{\langle k\rangle\lambda_{0}}\,\epsilon^{-l} ~.
\end{eqnarray}
This nontrivial solution vanishes at threshold values $\lambda_{0_{c}}$ 
in the usual form $s \sim \lambda_{0}-\lambda_{0_{c}}$, 
where the critical points $\lambda_{0_{c}}$ are given in terms of the parameters by  
\begin{eqnarray}\label{eq4}
\lambda_{0_{c}}=\frac{\alpha}{\langle k\rangle}\,\epsilon^{-l} ~.
\end{eqnarray}

%%%%%%%%%%%%%%%%%%%%%%%%%%%%%%%%%%%%%%%%%%%%%%%%%%%%%%%%%%%%%%%%%%%%%%%%%%
\begin{figure}[t]
\begin{center}
\vspace{0.5cm}
\includegraphics[width=0.47\textwidth,angle=0]{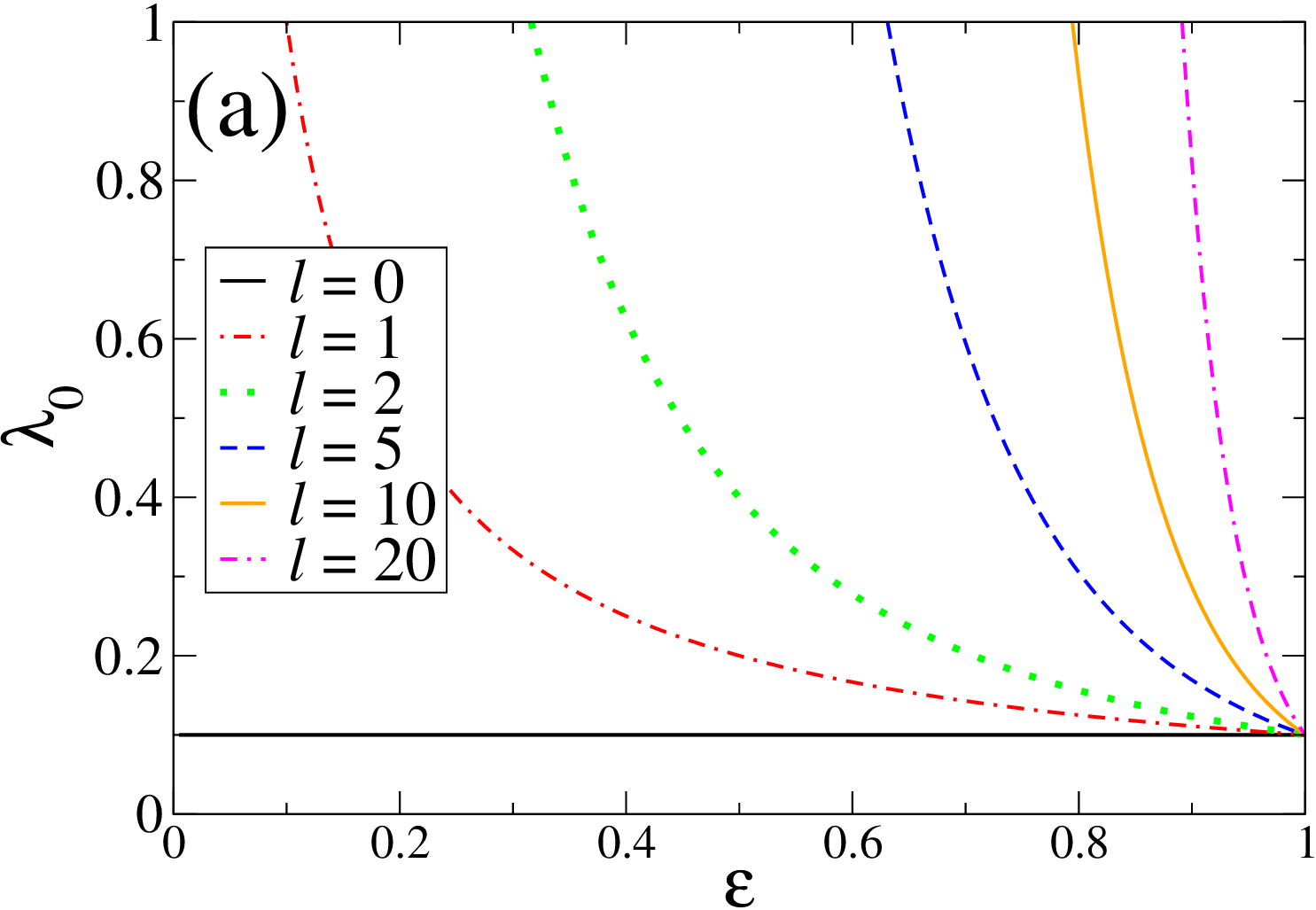} 
\includegraphics[width=0.47\textwidth,angle=0]{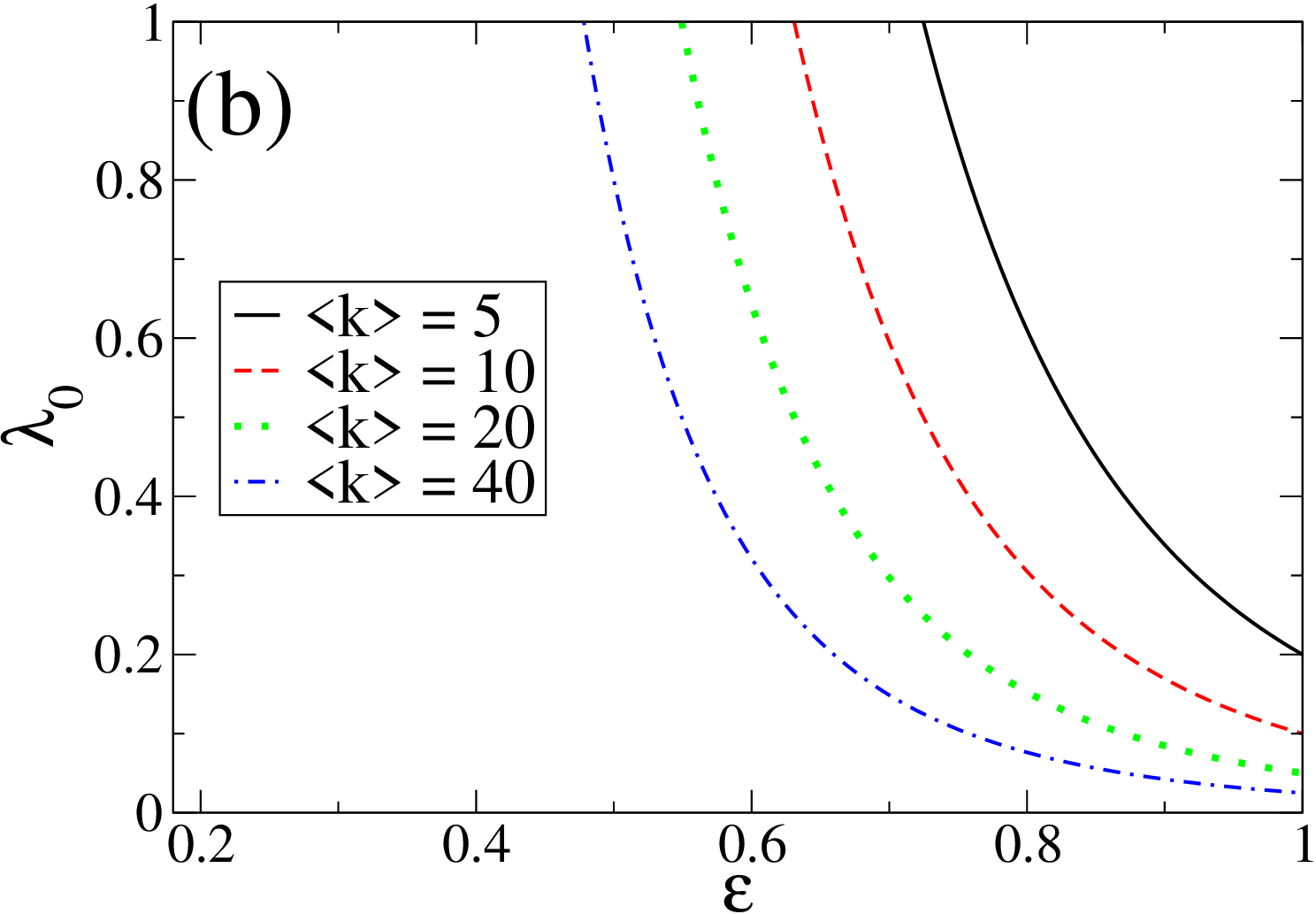}
\end{center}
\caption{(Color online) Phase diagram of the model, 
predicted by the mean-field approach, given by Eq. (\ref{eq4}), 
in the plane $\lambda_{0}$ versus $\epsilon$, for $\langle k\rangle=10$ 
and typical values of $l$ (a). 
The effect of varying  $\langle k\rangle$ is also exhibited for a given value of $l$, 
namely $l=5$ (b). 
The region above the curves represents the phase where information keeps being 
spread over the network.
}
\label{fig1}
\end{figure}
%%%%%%%%%%%%%%%%%%%%%%%%%%%%%%%%%%%%%%%%%%%%%%%%%%%%%%%%%%%%%%%%%%%%%%%%%%%

%%%%%%%%%%%%%%%%%%%%%%%%%%%%%%%%%%%%%%%%%%%%%%%%%%%%%%%%%%%%%%%%%%%%%%%%%%
\begin{figure}[t]
\begin{center}
\vspace{0.5cm}
\includegraphics[width=0.47\textwidth,angle=0]{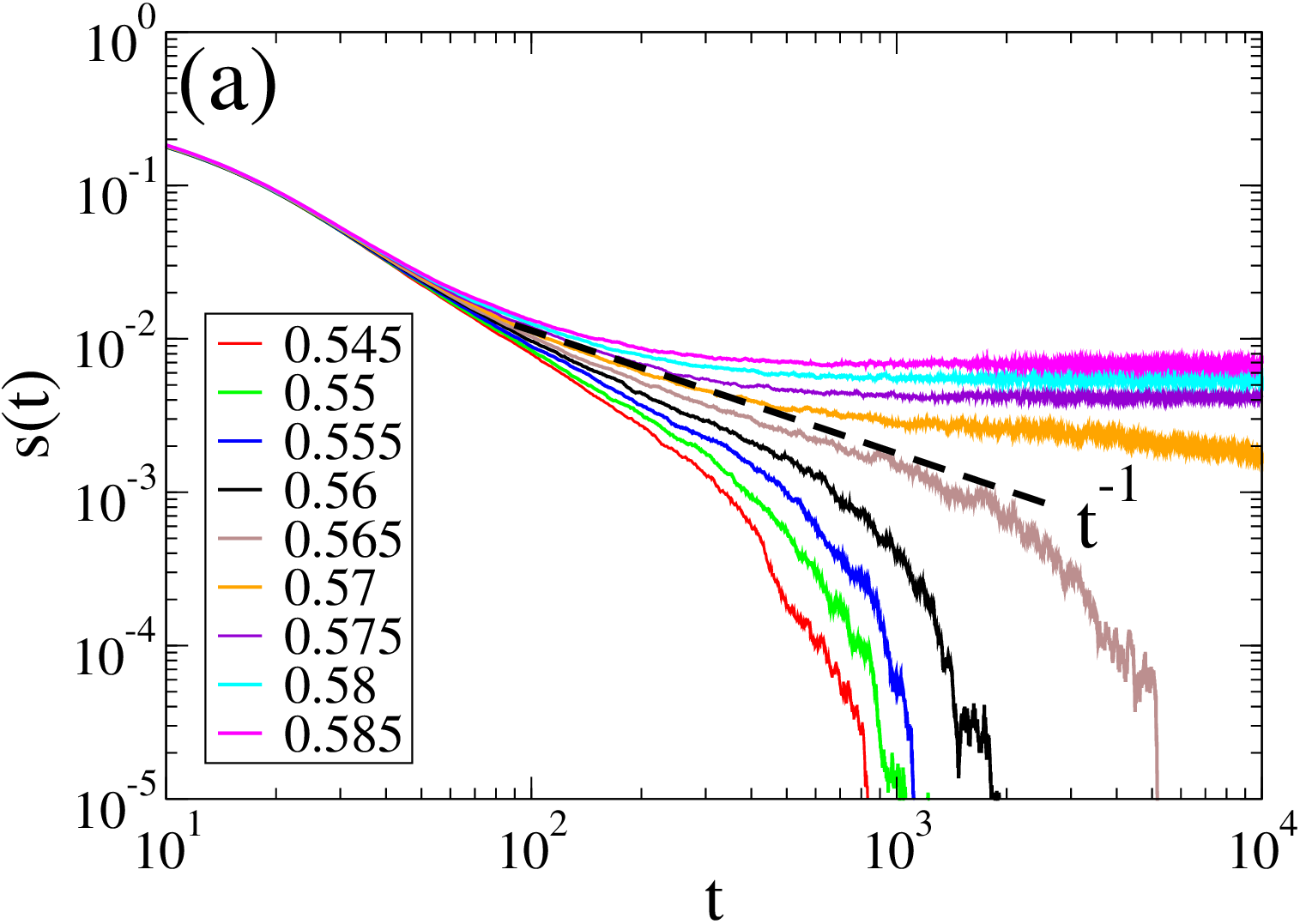}
\includegraphics[width=0.47\textwidth,angle=0]{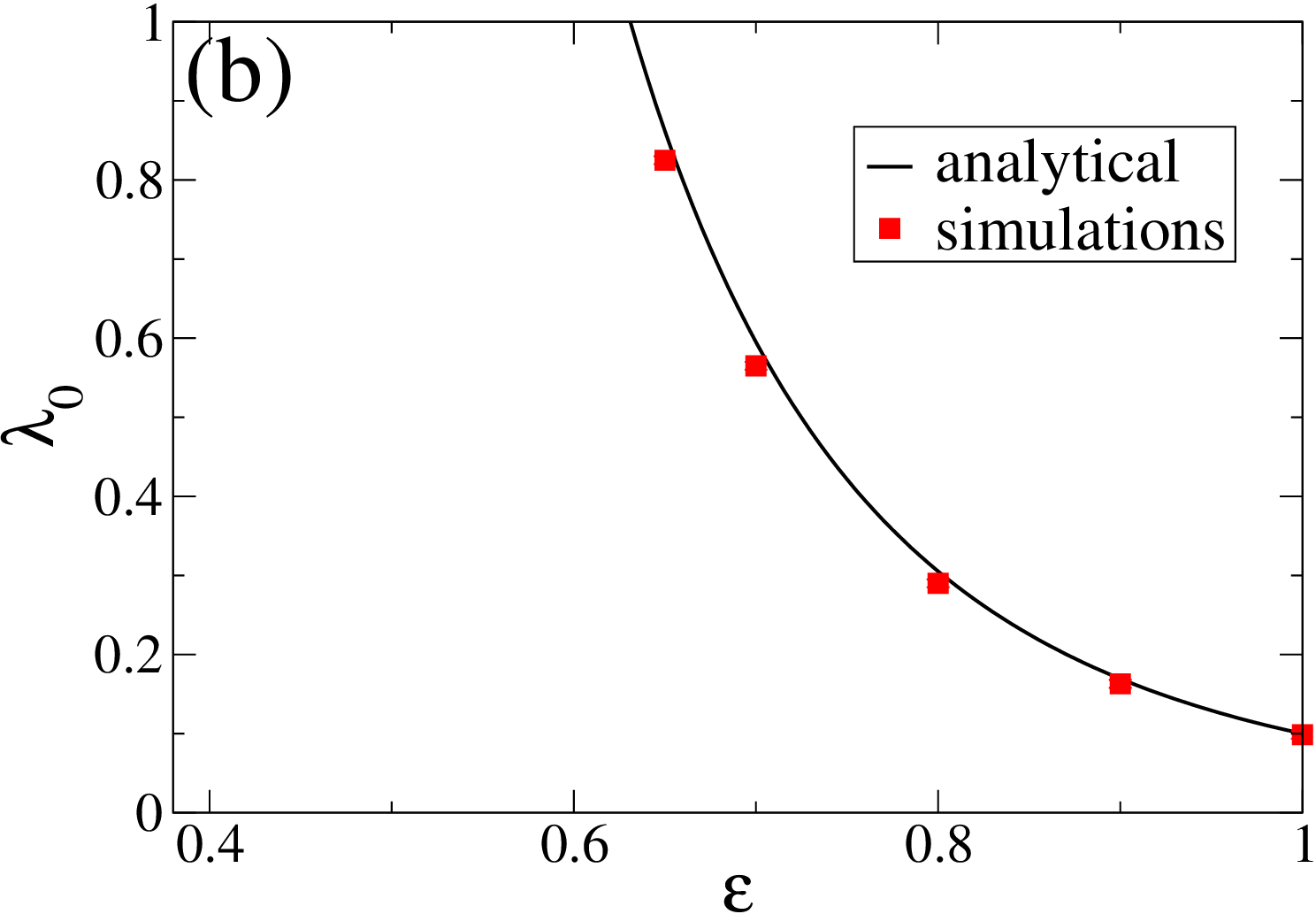}
\end{center}
\caption{(Color online) Time evolution of the density of Spreaders,  
$s(t)$,  in the vicinity of the transition, for typical values of $\lambda_{0}$ 
exhibited in the legend (a). 
Right at the critical point, the density of Spreaders decays as $s(t) \sim t^{-1}$, 
as indicated by the dashed line for $\lambda_{0}=0.565$. 
These results were obtained from simulations of the model with uniform $\epsilon$ and $l$, 
with $\langle k\rangle=10$, $l=5$, $\epsilon=0.7$, $N=10^{5}$ nodes and $100$ independent simulations. 
It is also exhibited the comparative phase diagram in the plane $\lambda_{0}$ versus $\epsilon$ 
for $l=5$ and $\langle k\rangle=10$ (b). 
The squares were obtained from the simulations, whereas the full line is the 
analytical prediction, given by Eq. (\ref{eq4}). Error bars are smaller than symbol size.}
\label{fig2}
\end{figure}
%%%%%%%%%%%%%%%%%%%%%%%%%%%%%%%%%%%%%%%%%%%%%%%%%%%%%%%%%%%%%%%%%%%%%%%%%%%
\noindent
In other words, for given values of the parameters $\epsilon$ and $l$ 
\cite{footnote2}, 
there is a critical value of the spreading probability $\lambda_{0_{c}}$ 
given by Eq. (\ref{eq4}) separating a phase where the information stops 
being spread (for $\lambda_{0}<\lambda_{0_{c}}$) and a phase where a 
certain fraction of the population remains spreading the information 
(for $\lambda_{0}>\lambda_{0_{c}}$). In Fig. \ref{fig1} (a) we exhibit 
the phase diagram of the model in the plane $\lambda_{0}$ versus $\epsilon$ 
for $\langle k\rangle=10$ and typical values of $l$. 
One can see that the larger $l$, the smaller the region where 
the information keeps spreading (the region above the curves). 
This result is easily understood: if we allow the spreading probabilities 
to decrease a large number of times, the probability of the Restrained 
individuals to become Spreaders becomes very small, and it is improbable that 
the information will remain being spread over the network. In fact, this event will 
occur only if the initial spreading probability $\lambda_{0}$ is large. 
It is also shown, in Fig. \ref{fig1} (b), the phase diagram for a fixed value 
of $l$ ($l=5$) and different values of the average degree $\langle k\rangle$. 
In this case, the spreading phase increases with $\langle k\rangle$. 
In fact, if each individual has on average a large number 
of contacts (nearest neighbors), there is a greater possibility of spreading 
the information across the network in comparison with the case of a small 
average degree.

We confronted the analytical solution of the model with numerical results. 
We have simulated the model on ER random graphs 
with $N=10^{5}$ nodes and different values of the parameters. 
For each value of $\epsilon$ and $l$, we have considered different values of 
$\lambda_{0}$ and $100$ independent simulations (furnishing configurational averages). 
As initial condition,  1\% of the network nodes were set in the $S$ state 
and the remaining ones in the $R$ state. 
The threshold values $\lambda_{0_{c}}$ were estimated from the time evolution of 
the density of Spreaders, $s(t)$. Right at the critical point $\lambda_{0_{c}}$, 
this quantity decays in time as the power law $s(t) \sim t^{-\delta}$  
at mean-field level \cite{hinrichsen,dickman}. As an example, we exhibit in Fig. \ref{fig2} (a) 
the time evolution of $s$ for $l=5$, $\epsilon=0.7$ and some values of $\lambda_{0}$ 
in the vicinity of the transition. One can see that the above-mentioned power-law 
behavior can be observed for $\lambda_{0}\sim 0.565$. We repeated this procedure 
for other values of $\epsilon$, and we compared the estimated values of the 
thresholds $\lambda_{0_{c}}$ with the values obtained from the mean-field approach, 
Eq. (\ref{eq4}). We can see from Fig. \ref{fig2} (b) that the considered size 
($N=10^{5}$ nodes) gives us a good estimate of the critical points $\lambda_{0_{c}}$, 
which confirms that the assumptions made to analytically solve the model are valid.

%%%%%%%%%%%%%%%%%%%%%%%%%%%%%%%%%%%%%%%

\subsection{Uniform $\epsilon$ and random $l$}

  In this case,   $\epsilon_{j}=\epsilon$ for all $j$ while  $l_{j}$ 
is different for each individual $j$. In other words, we have an additional 
heterogeneity in the population due to the individual capacity to decrease 
the spreading probability a distinct number of times, i.e., each agent $j$ 
has a limiting parameter $l_{j}$ that is an integer number generated from a 
uniform distribution in the range $[0,10]$.

%%%%%%%%%%%%%%%%%%%%%%%%%%%%%%%%%%%%%%%%%%%%%%%%%%%%%%%%%%%%%%%%%%%%%%%%%%
\begin{figure}[t]
\begin{center}
\vspace{0.3cm}
\includegraphics[width=0.48\textwidth,angle=0]{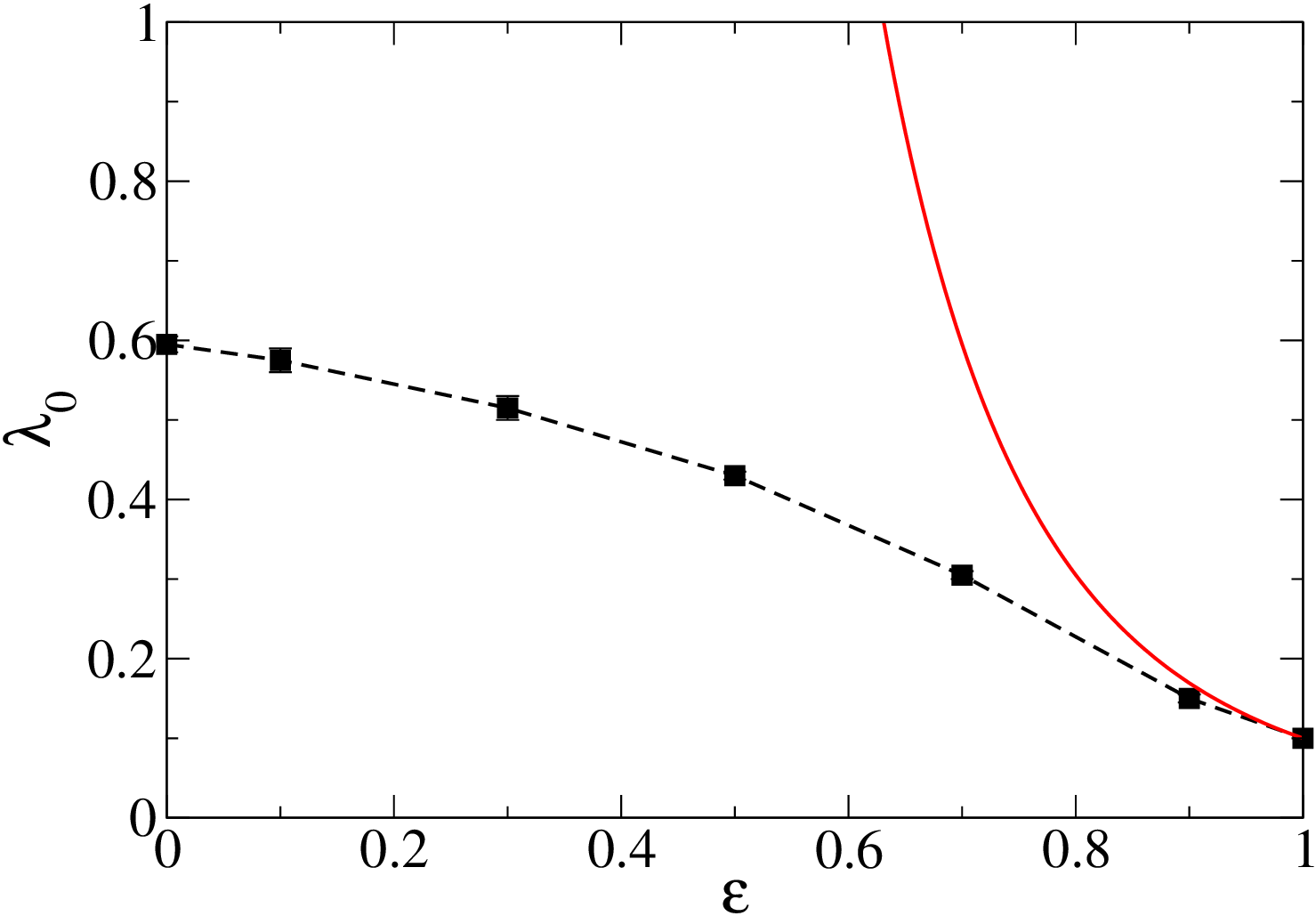}
\end{center}
\caption{(Color online) Phase diagram in the plane $\lambda_{0}$ versus $\epsilon$,  
for $\langle k\rangle=10$ and $l$ uniformly distributed in the integer range [0,10]. 
The squares were obtained from the simulations, whereas the dashed line is just 
a guide to the eyes. Error bars are smaller than symbols. 
For comparison, the frontier of the uniform case for $l=5$ 
is also exhibited (red full line).}
\label{fig3}
\end{figure}
%%%%%%%%%%%%%%%%%%%%%%%%%%%%%%%%%%%%%%%%%%%%%%%%%%%%%%%%%%%%%%%%%%%%%%%%%%%

Following the above-discussed procedure, we have analyzed the time evolution 
of the density $s(t)$ of Spreaders for populations of size $N=10^{5}$. 
The critical points $\lambda_{0_{c}}$ were estimated from the power-law behavior 
$s(t)\sim t^{-\delta}$ as in the previous case in Section III.A. 
The critical line  in the plane $\lambda_{0}$ 
versus $\epsilon$, for  $l$ uniformly distributed in [0,10] 
is exhibited in Fig. \ref{fig3} (points joined by a dashed line). 
In contrast to what happens in the uniform case presented in Section III.A, 
when  $l$ is random the relation between $\lambda_{0_{c}}$ and $\epsilon$ is not a power law 
(see the comparison in Fig. \ref{fig3}). 
As a consequence of the heterogeneity of $l$, 
one can observe that there is a phase transition even for $\epsilon=0$, 
which turns the spreading phase (above the dashed curve in Fig. \ref{fig3}) 
larger than in the case of uniform values of $l$ and $\epsilon$ 
(for comparison, we also exhibit in Fig. \ref{fig3} 
(full line) the frontier of the uniform case for $l=5$, corresponding to the 
mean value of $l_j$). 
In other words, the transition is not eliminated even for $\epsilon=0$. 
This fact can be understood as follows. In the case of a small value of $\epsilon$, 
the spreading probabilities decrease fast, but there are some individuals 
for which this decrease  occurs a small number of times  (e.g. for $l=1$)
or does not occur at all (for $l=0$). 
Thus, if we consider simulations with a large initial value 
of $\lambda$, namely $\lambda_{0}\gtrsim 0.6$, those individuals are responsible 
by the permanent spread of the information through the network. 
Notice that in the limiting case $\epsilon=0$, the individuals $j$ 
with $l_{j}=0$ (that are around $10\%$ of the population) keep their 
spreading rates equal to the initial value, i.e., they have 
$\lambda_{j}=\lambda_{0}$ at all time steps. These individuals can 
spread permanently the information across the network if the initial 
value of the spreading probability is larger than $\approx 0.6$ (see Fig. \ref{fig3}). 
In sum, for distributions with a given mean value of $l$, the enhancement of the 
transmission region above the curves is more pronounced if  $l=0$ is allowed and 
the larger is the dispersion.

%%%%%%%%%%%%%%%%%%%%%%%%%%%%%%%%%%%%%%%

\subsection{Random $\epsilon$ and uniform $l$}

%%%%%%%%%%%%%%%%%%%%%%%%%%%%%%%%%%%%%%%%%%%%%%%%%%%%%%%%%%%%%%%%%%%%%%%%%%
\begin{figure}[t]
\begin{center}
\vspace{0.3cm}
\includegraphics[width=0.48\textwidth,angle=0]{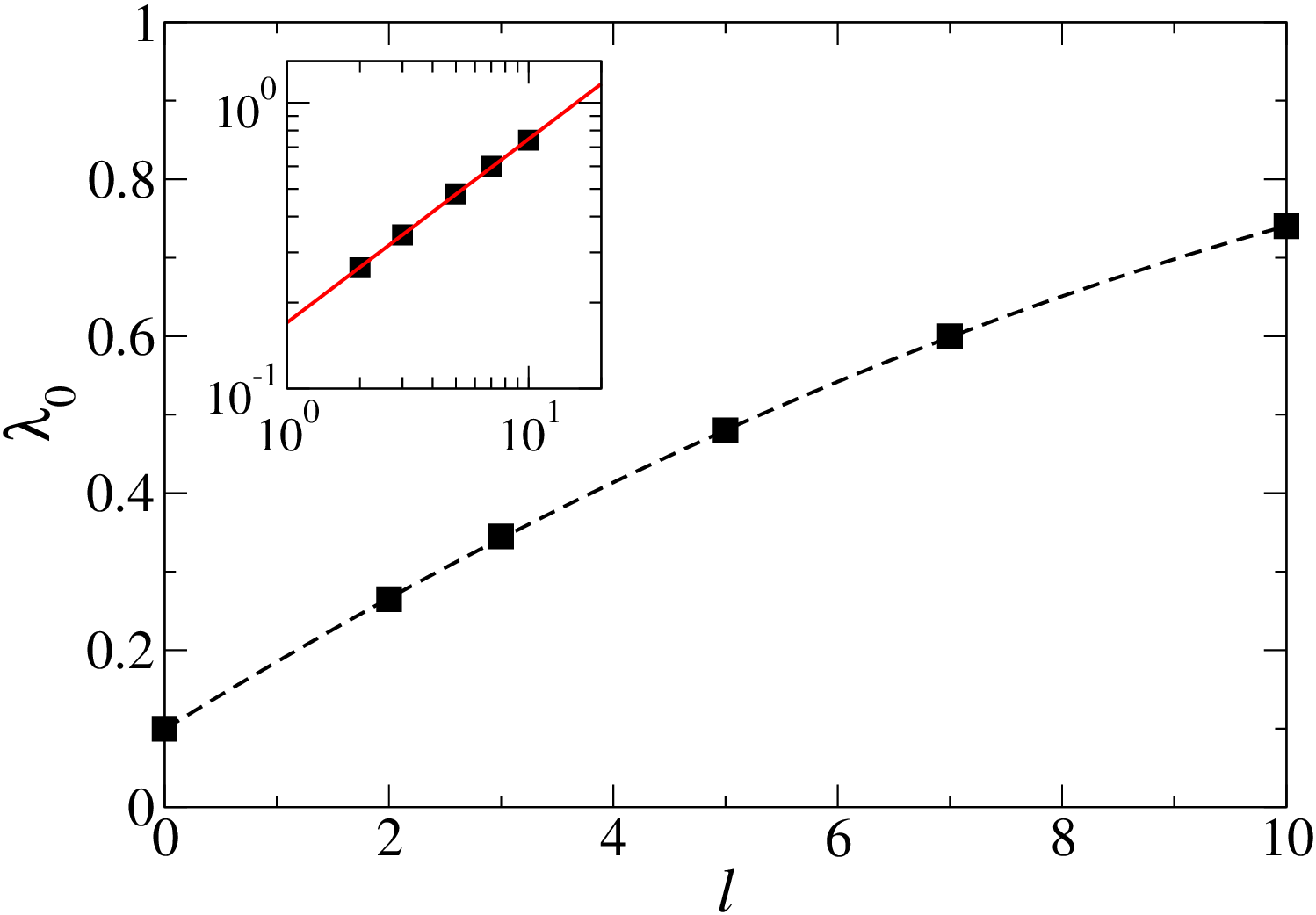}
\end{center}
\caption{(Color online) Phase diagram in the plane $\lambda_{0}$ versus $l$,  
for $\langle k\rangle=10$ and $\epsilon$ uniformly distributed in the range [0,1]. 
The squares were obtained from the simulations, whereas the dashed line is just 
a guide to the eyes. Error bars are smaller than symbols. 
The inset shows the same plot in log-log scale. 
Fitting data, we obtained the power-law relation $\lambda_{0_{c}}\sim l^\gamma$, 
with $\gamma \simeq {0.64}$ (red full line). 
}
\label{fig4}
\end{figure}
%%%%%%%%%%%%%%%%%%%%%%%%%%%%%%%%%%%%%%%%%%%%%%%%%%%%%%%%%%%%%%%%%%%%%%%%%%%

  In this case, we have $l_{j}=l$ while $\epsilon$ is different for 
each individual $j$. In other words, we have an additional heterogeneity in 
the population due to the individual rate of decrease of the spreading 
probability, i.e., each agent $j$ has a decreasing rate $\epsilon_{j}$ 
that is a real number generated from a uniform distribution in the range $[0,1]$.

Again, we have followed the usual procedure and we have analyzed the time evolution 
of the density of spreaders $s(t)$. In the present formulation, as $\epsilon$ is random, 
we have plotted the phase diagram in the plane $\lambda_{0}$ versus $l$. 
One can observe in Fig. \ref{fig4} that the region above the curve, 
for which the information is continuously spread across the network, 
decreases for increasing values of $l$. In fact, 
if we increase $l$, the final spreading rates $\lambda_{l}=\lambda_{0}\,\epsilon^{l}$ 
are small even if $\epsilon>>0$, and in this case it is difficult for the Restrained 
individuals to become Spreaders again. If we plot the data in the log-log scale, 
one case see that the quantities are related by the power law 
$\lambda_{0_{c}}\sim l^\gamma$, with $\gamma \simeq {0.64}$   
(see the inset of Fig. \ref{fig4}).

%%%%%%%%%%%%%%%%%

\subsection{Random $\epsilon$ and random $l$}

In this case, we have the more heterogeneous instance where both parameters are random: 
$\epsilon$ is uniformly distributed in the range $[0,1]$ and $l$ 
is an integer number uniformly distributed in the range $[0,10]$. 
Although in this case there is no phase diagram to plot, 
we can discuss the criticality of the model at the single transition point 
$\lambda_{0_{c}}$. Our numerical estimate is $\lambda_{0_{c}} = 0.31 \pm 0.005$. 
In other words, 
for $\lambda_{0}>0.31$ the information is permanently disseminated through the network 
by a finite fraction of the population. 
Notice that the threshold  is relatively small. Thus, the enhanced diversity of spreading 
probabilities $\lambda_{j}$ in the population favors the propagation of information.

% ##################################################################

\section{Final Remarks}

  In this work we have studied a model of information spreading on complex networks. 
The model is based on the SIS epidemic model, but the ``infection'' or spreading 
rates vary with time, decaying with the number of ``reinfections''. 
This decay is controlled by two parameters, $\epsilon$ and $l$, and we have considered 
that they can be either uniform or random. These features make the population heterogeneous, 
since the agents may have distinct rates of transition between the two possible states, 
namely, Spreader (S) or Restrained (R).

We solved the model analytically in its simplest formulation, for constant $\epsilon$ 
and $l$. In this case, the critical spreading rates are given by 
$\lambda_{0_{c}}=(\alpha/\langle k\rangle)\,\epsilon^{-l}$, where $\lambda_{0}$ 
is the initial probability (at $t=0$) of the transition $R\to S$, $\alpha$ 
is the probability of the transition $S\to R$ and $\langle k\rangle$ is the average 
degree of the random network. 
The critical rates define frontiers 
(for different values of $l$) in the plane $\lambda_{0}$ versus $\epsilon$ that 
separate a phase where the information stops being spread 
(for $\lambda_{0}<\lambda_{0_{c}}$) and a phase where a certain fraction of 
the population remains spreading the information (for $\lambda_{0}>\lambda_{0_{c}}$). 
This result suggests that networks with large average degree favor the spreading 
of information, as well as populations with a small capacity to decrease 
the spreading rates (i.e., small $l$), as intuitively expected. 
All analytical results were confirmed by Monte Carlo simulations.

In the case where one (or both) of the parameters $\epsilon$ or $l$ is (are) random, 
the model was analyzed only through numerical simulations. 
For random $l$ and fixed $\epsilon$, we have observed that information spreading  
is favored and thus the spreading phase, where the information is permanently 
disseminated through the network by a finite fraction of the population, 
is larger than in the uniform case. For random $\epsilon$ and fixed 
$l$ the two phases are of comparable size, 
and the power law $\lambda_{0_{c}}\sim l^\gamma$, with $\gamma \simeq {0.64}$ arises. 
Finally, in the case of both parameters being heterogeneous in the population, 
we have found the transition at $\lambda_{0_{c}}\simeq 0.31$. Thus, one can 
conclude that the more heterogeneous the population is, more the information 
spreading is favored. In other words, population diversity is an interesting feature 
to be taken into account in models of information/rumor spreading. 

Let us remark that, due to the correspondence between the present model and 
the SIS model with aging, our results can be immediately applied to the 
latter model. In that case, diversity would be malefic, 
since disease propagation will be favored.

\section*{Acknowledgements}

The authors are grateful to Jose Fernando Mendes for having provided the 
computational resources of the Group of Complex Systems and Random Networks 
(GNET) of the Aveiro University, Portugal, where the simulations were performed. 
This work was supported by the Brazilian funding agencies FAPERJ,
CAPES and CNPq.


\begin{thebibliography}{40}



\bibitem{loreto_rmp}
C. Castellano, S. Fortunato, V. Loreto, Rev. Mod. Phys. \textbf{81}, 591 (2009).


% modelos:

\bibitem{dk} %daley-kendall
D. J. Daley, D. G. Kendall, Nature \textbf{204}, 1118 (1964).

\bibitem{galam}
S. Galam, Physica A \textbf{320}, 571 (2003).


\bibitem{moreno1}
M. Nekovee, Y. Moreno, G. Bianconi, M. Marsili, Physica A \textbf{374}, 457 (2007).

\bibitem{moreno2}
Y. Moreno, M. Nekovee, A. F. Pacheco, Phys. Rev. E \textbf{69}, 066130 (2004).


\bibitem{vazquez}
A. Vazquez, Phys. Rev. E \textbf{74}, 056101 (2006).

\bibitem{zanette}
D. Zanette, Phys. Rev. E \textbf{64}, 050901(R) (2001).

\bibitem{tang}
D. Trpevski, W. K. S. Tang, L. Kocarev, Phys. Rev. E \textbf{81}, 056102 (2010).

\bibitem{zhao1} %SIHR rumor
L. Zhao,  J. Wang, Y. Chen, Q. Wang, J. Cheng,  H. Cui, Physica A \textbf{391}, 2444 (2012).

\bibitem{zhao2} %effects of media
L. Zhao, Q. Wang, J. Cheng, D. Zhang, T. Ma, Y. Chen, J. Wang, Physica A \textbf{391}, 3978 (2012).

\bibitem{zhao3} %forgetting mechanism
L. Zhao, Q. Wang, J. Cheng, Y. Chen, J. Wang, W. Huang, Physica A \textbf{390}, 2619 (2011).

\bibitem{zhao4} %forgetting and remebering mechanisms
L. Zhao,  X. Qiu, X. Wang, J. Wang, Physica A (2012), doi:10.1016/j.physa.2012.10.031.

\bibitem{moro}
J. L. Iribarren, E. Moro, Phys. Rev. Lett. \textbf{103}, 038702 (2009).

\bibitem{kun}
G. Kocsis, F. Kun, J. Stat. Mech. P10014 (2008).


\bibitem{anderson_may}
R. M. Anderson, R. M. May, \textit{Infectious Diseases of Humans: Dynamics and Control} 
(Oxford University Press, Oxford, 1991).



\bibitem{toole} % ``adoption of inonvation''
J. L. Toole, M. Cha, M. C. Gonz\'alez, PLoS ONE 7(1):e29528 (2011).

\bibitem{iglesias} % tambem sobre ``adoption of inonvation''
S. Gon\c{c}alves, M. F. Laguna, J. R. Iglesias, Eur. Phys. J. B \textbf{85}, 192 (2012).

\bibitem{er}
P. Erd\H{o}s, A. R\'enyi, Publ. Math. Debrecen \textbf{6}, 290  (1959).


\bibitem{pastor_satorras}
R. Pastor-Satorras, A. Vespignani, Phys. Rev. E \textbf{63}, 066117 (2001).

\bibitem{footnote1}
In the following, we will verify that this assumption is valid.


\bibitem{nuno_marcio}
N. Crokidakis, M. A. de Menezes, J. Stat. Mech P05012 (2012).

\bibitem{footnote2}
Without loss of generality, we have considered $\alpha=1$.

\bibitem{hinrichsen}
H. Hinrichsen, Adv. Phys. \textbf{49}, 815 (2000).


\bibitem{dickman}
J. Marro, R. Dickman, \textit{Nonequilibrium Phase Transitions in Lattice Models} 
(Cambridge University Press, Cambridge, 1999).


\end{thebibliography}
\end{document}